\documentstyle[12pt,epsf,cite]{article}
\begin{document}
\title{Study of supernarrow dibaryon production in process
$pd\to pX$}
\author
{L.V.~Fil'kov$^a$, V.L.~Kashevarov$^a$, E.S.~Konobeevskiy$^b$,\\ 
M.V.~Mordovskoy$^b$, 
S.I.~Potashev$^b$, 
V.M.~Skorkin$^b$\\
$a$ -- {\it Lebedev Physical Institute, Moscow, Russia} \\
$b$ -- {\it Institute of Nuclear Research, Moscow, Russia}}
\maketitle

\begin{abstract}
The reactions $p+d\to p+p(\gamma n)$ and $p+d\to p+ d(\gamma)$ 
at 305 MeV are studied with the aim to search for supernarrow 
dibaryons. The experiments were carried out at the Moscow Meson 
Factory using a spectrometer TAMS, which detected two charged 
particles at various angles. Narrow structures in missing-mass 
spectra at 1905 and 1924 MeV have been observed. An analysis of 
the angular dependence of the experimental data shows that the 
resonance at $M$=1905 MeV most likely corresponds to the 
production of the isovector supernarrow dibaryon.  
\end{abstract}

Recently, a number of works have appeared in which narrow dibaryons
were searched for in experiments on collisions of the nucleon and a
few-body system at intermediate energies \cite{kom}. 

Early, we carried out measurements of missing-mass spectra in
the reactions $pd\to pX\to pd(\gamma)$ and $pd\to pX\to pp(n\gamma)$
at an incident proton energy of 305 MeV \cite{filk1} using a 
double-arm spectrometer. The narrow structure in the missing-mass 
spectrum at 1905 MeV with a width equal to the experimental 
resolution 7 MeV was observed in this experiment.

Now, we present further results of the study of the
reactions under consideration with the improved facility. The 
experiments were performed with the proton accelerator of the Moscow
Meson Factory at 305 MeV. A proton beam alternately bombarded
CD$_2$ and $^{12}$C targets. 
The $pd$-reaction contribution was determined 
by subtraction of the $^{12}$C spectrum from the CD$_2$ one. 
The two-arm spectrometer TAMS detected the scattered proton  
in coincidence with the second charged particle ($p$ or $d$).

The left movable spectrometer arm, being a single telescope 
$\Delta E-\Delta E-E$, was used to measure the energy and time 
of flight of the scattered proton at $\theta_L=72.5^{\circ}$ 
(or $70^{\circ}$ in another run). The right fixed arm detected 
the proton or the deuteron from the expected dibaryon decay. 
It consisted 
of three telescopes, which were located at $\theta_R=33^{\circ}$,
$35^{\circ}$, and $37^{\circ}$. These angles correspond to the 
directions of motion of the produced dibaryons with the chosen mass 
ranges. Each telescope included a full absorption detector 
and two thin plastic $\Delta E$ detectors for a time-of-flight 
measurement. A trigger was generated by a coincidence of the 
$\Delta E$ detector signals of the left arm with those
of any right-arm telescope. Selected by a coincidence, the $E$-signals
of the scattered proton form its energy spectrum and, accordingly, a 
missing-mass spectrum. The spectrometer was calibrated using 
the peak of elastic $pd$ scattering  \cite{ald}.

The experimental missing-mass spectra obtained on the targets of
deuteried polyethylene and carbon are shown in Figs. $1a-1c$. Each
spectrum corresponds to a certain combination of outgoing angles
of the scattered proton and the second charged particle. 
These combinations in Figs. $1b$ and $1c$ are consistent with the 
change in the emission angle $\theta_R$ of a dibaryon with the given 
mass when the angle $\theta_L$ is equal to $70^{\circ}$ or 
$72,5^{\circ}$. As evident from Fig. 1, resonance-like 
behavior of the spectra is observed in two mass regions for the 
CD$_2$ target, while the spectra for the carbon target are smooth 
enough \cite{izv}.

There are $58\pm 13$ events in the peak at $1905\pm 2$ MeV 
which is shown in Fig. $1b$. 
The statistical significance of this 
resonance is 4.5 standard deviations. 
The spectrum in Fig. $1a$ shows the other peak with the mass 
$M=1924\pm 2$ MeV containing $79\pm 16$ events. 
The statistical significance of this structure is 4.7 S.D.
The widths of both observed peaks correspond to the experimental
resolution (3 MeV). 
The peak at 1924 MeV was only obtained for one spectrum close 
to the upper limit of the missing mass. In the other cases, this mass 
position was out of the range of measurement. 
Therefore, in the present work, we restrict ourselves only to an 
analysis of the peak at 1905 MeV.

The experimental missing-mass spectra in the range of
1895--1913 MeV, after
subtracting the carbon contributions, are shown in Figs. $2a-2c$.

As seen from Figs. 1 and 2, the resonance behaviour of the cross 
section exhibits itself in a limited angular region. 

If the observed structure at $M=1905$ MeV corresponds to a
dibaryon decayed mainly into two nucleons, then the expected
angular cone size of the emitted nucleons would be about 
50$^{\circ}$. Moreover, the angular distributions of the emitted 
nucleons are expected to be very smooth in the angle region under 
consideration. Thus, even assuming the dibaryon production cross 
section to be equal to an elastic scattering one (40 $\mu$b/sr), 
their contribution to the missing-mass spectra in Fig. $2a-2c$ 
would be nearly the same and would not exceed a few events. 
Hence, the found peaks are hardly interpreted as a manifestation of
the formation and the decay of such states.

In \cite{filk2,ak,ger,al1} supernarrow dibaryons were 
considered, whose decay into two nucleons is suppressed 
by the Pauli exclusion principle. Such states with a mass
$M<2m_N + m_{\pi}$ ($m_N$ and $m_{\pi}$ are the masses of the 
nucleon and the pion) can decay mainly with a photon emission.
  
Using the Monte Carlo simulation, we estimated the contribution 
of the supernarrow dibaryons with different quantum numbers and 
$M$=1905 MeV to the mass spectra at various angles of the 
left and right arms of our setup. 
The production cross section and branching ratio of these 
states were taken from \cite{filk1}. 
The obtained results are listed in the table.  
\begin{table}
\caption{The expected contributions of the supernarrow dibaryons into
the mass spectra at various angles of the left and right arms of 
the setup.}
\begin{center}
\begin{tabular}{|l|r|r|r|r|r|r|r|r|r|r|} \hline
\multicolumn{1}{|c|}{$\theta_L$}&\multicolumn{5}{|c|}{$70^{\circ}$}&
\multicolumn{5}{|c|}{$72.5^{\circ}$} \\ \hline
\multicolumn{1}{|c|}{$\theta_R$}&$31^{\circ}$&$33^{\circ}$&$35^{\circ}$
&$37^{\circ}$&$39^{\circ}$&$31^{\circ}$&$33^{\circ}$&$35^{\circ}$
&$37^{\circ}$&$39^{\circ}$ \\ \hline
\multicolumn{1}{|l|}{$T,J^P$}&\multicolumn{10}{|c|}{Dibaryon mass
$M=1905\pm 2$ MeV} \\ \hline
$0,0^+\;\gamma pn$ & 3&  2&  3&  2&  2&  3&  3&  3&  2& 1 \\ \hline
$0,0^+\;\gamma d $ & 0& 11&  5& 10&  0&  1&  7& 10&  0& 0 \\ \hline
$0,0^-\;\gamma pn$ & 1&  1&  1&  1&  1&  1&  2&  1&  1& 1 \\ \hline
$0,0^-\;\gamma d $ & 0&  6&  3&  5&  0&  1&  4&  5&  0& 0 \\ \hline
$1,1^+\;\gamma pn$ & 5& 16& 24& 15&  4& 14& 29& 24&  7& 2 \\ \hline
$1,1^-\;\gamma pn$ & 9& 36& 53& 33&  8& 31& 64& 52& 15& 4 \\ \hline
\end{tabular}
\end{center}
\end{table}

This calculation showed that the angular cone of charged particles 
emitted from a certain dibaryonic state can be narrow enough. 
An axe of this cone is lined up with the direction of the dibaryon 
emission. Therefore,  by placing the right spectrometer arm at an 
expected angle of the dibaryon emission, we essentially increase 
the signal-to-background ratio.

 In Fig. $2a-2c$, the experimental spectra are
compared with the predicted yields normalized to the maximum of
the measured signal in Fig. $2b$. The solid and dashed curves in 
this figure correspond to states with 
isospin $T$=1 and $T$=0, respectively. 
  
As seen from this figure and the table, the ratios of the
calculated contributions to the given spectra are expected to be 
$0.3:1: 0.7$ if  the state at 1905 MeV is interpreted as an isovector
dibaryon [$D(T=1,J^P=1^{+})$ or $D(1,1^-)$].
This is in agreement with our experimental data within the
errors. On the contrary, the signals from isoscalar
dibaryons [$D(0,0^+)$ or $D(0,0^-)$] could be 
observed in Figs. $2b$ and $2c$ with the same probability. 

The following conclusions could be drawn : 1) as a result of the study
of the reactions $pd\to pd(\gamma)$ and $pd\to pp(\gamma n)$, 
two narrow structures at 1905 and 1924 MeV with widths of
less than 3 MeV were observed in the missing-mass spectra; 
2) the analysis of the angular dependence of the
experimental and theoretical yields of the reactions under 
consideration showed that the found peak at 1905 MeV can be 
explained as the manifestation of the supernarrow
dibaryons, the decay of which into two nucleons is suppressed by
the Pauli exclusion principle; 3) it is most likely that the 
observed state has a isospin equal to 1.

\newpage

\begin{figure}[htp]
\epsfxsize=16cm
\epsfysize=12cm
\epsffile{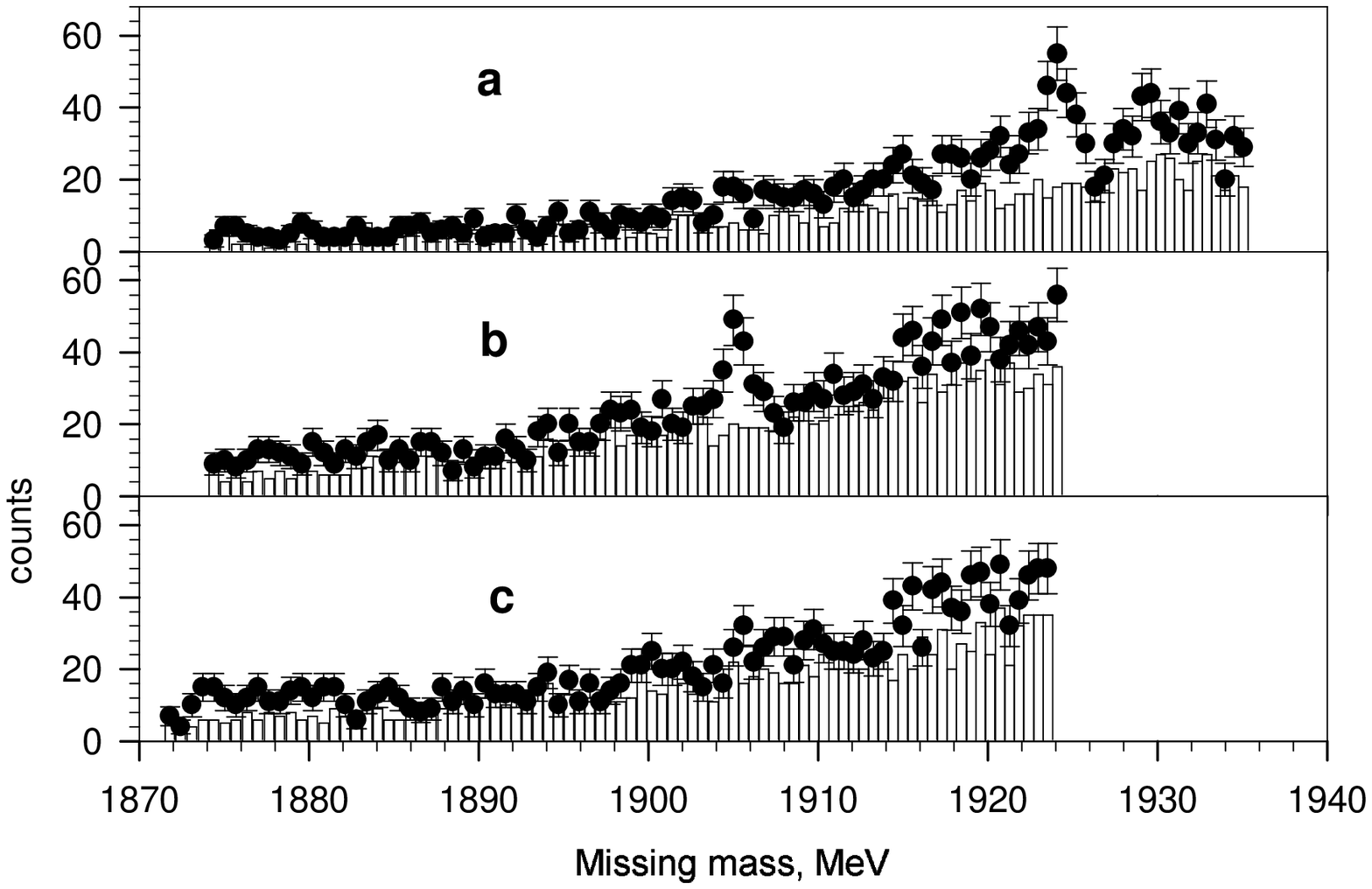}
\caption{ Missing-mass spectra: 
a) from the run data at 
$\theta_L=70^{\circ}$ and $\theta_R=33^{\circ}$, 
b) from the run data at $70^{\circ}$
and $35^{\circ}$, summarized with the run 
data at $72.5^{\circ}$ and $33^{\circ}$, 
c) from the run data at $70^{\circ}$ 
and $37^{\circ}$, summarized with the run data at 
$72.5^{\circ}$ and $35^{\circ}$. 
The points with the statistical errors correspond to
the experimental data obtained with a deuteried polyethylene and
the bars are ones obtained with  carbon as a target.}
\end{figure}
\newpage
\begin{figure}[htp]
\epsfxsize=16cm
\epsfysize=12cm
\epsffile{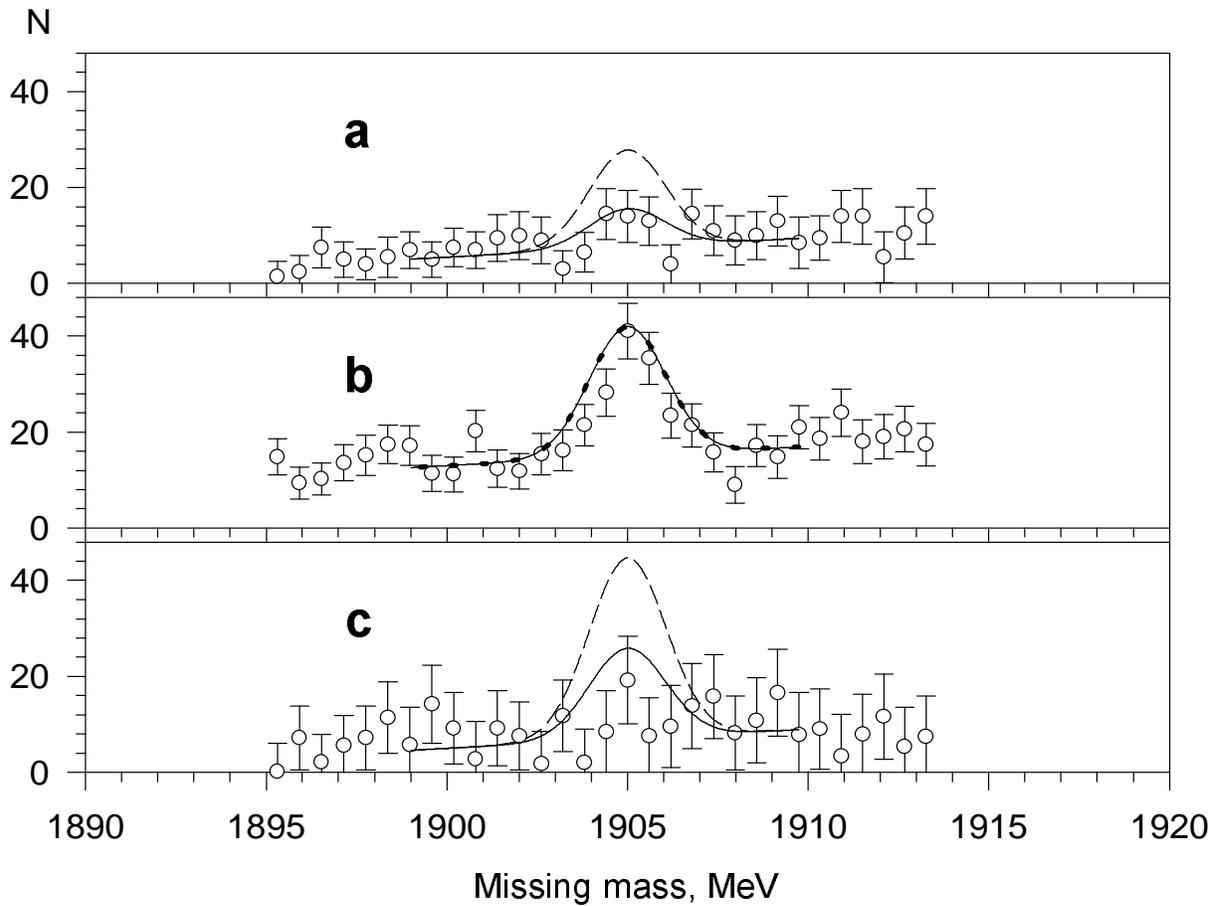}
\caption{ Missing mass spectra of the reaction on deuteron.
Areas marked by letters ($a$), ($b$), and ($c$) correspond to 
the same
experiment conditions as in Fig. 1. The solid and dashed curvers
are the theoretical prediction of the yields of the supernarrow
dibaryon with an isospin equal to 1 and 0, respectively.}
\end{figure}

\end{document}